\documentclass[11pt]{article}
\usepackage{moriond,epsfig}

\def\Journal#1#2#3#4{{#1} {\bf #2}, #3 (#4)}

\def\NPB{{\em Nucl. Phys.} B}
\def\PLB{{\em Phys. Lett.}  B}
\def\PRL{\em Phys. Rev. Lett.}
\def\PRD{{\em Phys. Rev.} D}

\def\APJ{\em Astrophys. J.}
\def\SNP{\em Sov. Jour. Nucl. Phys.}

\def\be{\begin{equation}}
\def\ee{\end{equation}}
\def\bea{\begin{eqnarray}}
\def\eea{\end{eqnarray}}

\begin{document}
\vspace*{4cm}
\title{Five Years of Neutrino Physics with Super-Kamiokande}

\author{ M.B. Smy\\ for the Super-Kamiokande Collaboration}

\address{Department of Physics and Astronomy, 4182 Frederick Reines Hall,
University of California, Irvine, Irvine, California 92697-4575, USA}

\maketitle\abstracts{
Using data from both solar and atmospheric neutrinos, Super-Kamiokande has
provided fundamental information on neutrino flavor mixing and
neutrino mass square differences.
}

\section{Introduction}

Both solar and atmospheric neutrino measurements have yielded
significantly smaller interaction rates than expected: the reduction
of the solar $\nu_e$ flux compared to the
standard solar model (SSM~\cite{ssm}) was observed to be
between 50\%~\cite{kam,gallex,sage,sk} and 65\%~\cite{cl,sno},
the atmospheric $\nu_e$ flux
was found to be the same within uncertainty while the atmospheric
$\nu_\mu$ flux was measured at about 65\% of the
calculation~\cite{skevidence}. Both deficits --- known as the solar
neutrino problem and the atmospheric anomaly --- are explained by
neutrino flavor changes caused by oscillations: solar neutrino experiments
detect predominantly $\nu_e$ and atmospheric experiments
are less sensitive to $\nu_\tau$ interactions.

Neutrino oscillations require the physical mass eigenstates
$\nu_i$ (with masses $m_1<m_2<$...) to differ from the flavor
eigenstates $\nu_\alpha$ (defined by charged-current interactions
with a charged lepton of flavor
$\alpha$). The flavor eigenstates $\nu_\alpha$ are orthogonal
linear combinations
of the mass eigenstates $\nu_i$: $\nu_\alpha=\sum_i U_{\alpha i}\nu_i$.
U is the unitary mixing matrix.
Since there are three charged lepton flavors ($e$, $\mu$ and $\tau$), 
only three neutrino flavor eigenstates can produce charged leptons via
charged-current interactions. The invisible width of the $Z^0$ boson is
compatible with decay into three light neutrinos, so only
three neutrino flavor eigenstates participate in neutral-current
interactions. If there are more than three neutrinos (and therefore
more than three neutrino flavor eigenstates), then some of the
flavor eigenstates are ``sterile'', that is, they feel neither
charged- nor neutral-current interactions.
In case of just three neutrinos (and neglecting CP violating phases),
the mixing matrix U can be described phenomenologically using the
three mixing angles $\theta_{12}, \theta_{13}$ and $\theta_{23}$.
Using the short notation
$s_{ij}=\sin\theta_{ij}, c_{ij}=\cos\theta_{ij}$ the mixing
matrix becomes
\[
U=\pmatrix{
 c_{12}c_{13}                    &  s_{12}c_{13}                    & s_{13} \cr
-s_{12}c_{23}-c_{12}s_{23}s_{13} &  c_{12}c_{23}-s_{12}s_{23}s_{13} & s_{23}c_{13} \cr
 s_{12}s_{23}-c_{12}c_{23}s_{13} & -c_{12}s_{23}-s_{12}c_{23}s_{13} & c_{23}c_{13}
}
\]
and the flavor conversion probability for a neutrino of energy $E$
after flight distance $L$ in vacuum is
\[
P_{\mbox{vac}}(\nu_\alpha\longrightarrow\nu_\beta)=
\delta_{\alpha\beta}-4\cdot\sum_{j<k} U_{\alpha j}U_{\beta j}U_{\alpha k}U_{\beta k}
\sin^2\frac{\Delta m^2_{jk}L}{4E}
\]
Atmospheric neutrino data requires~\cite{skevidence} a
$\Delta m^2_{\mbox{atm}}$
around $10^{-3}$eV$^2$, while solar neutrino data implies a
$\Delta m^2_{\mbox{sol}}$ below $\approx10^{-4}$eV$^2\ll\Delta m^2_{\mbox{atm}}$.
For the purpose of
this paper, the $\Delta m^2_{\mbox{sol}}$ is assumed to be $\Delta m^2_{12}$.
Then, $\Delta m^2_{\mbox{atm}}\approx\Delta m^2_{13}\approx\Delta m^2_{23}$ and
the atmospheric conversion probabilities are
\begin{eqnarray}
P_{\mbox{vac}}(\nu_e\longrightarrow \nu_\mu)&\approx&
\sin^22\theta_{13}\sin^2\theta_{23}\sin^2\frac{\Delta m^2_{\mbox{atm}}L}{4E}\cr
P_{\mbox{vac}}(\nu_\mu\longrightarrow \nu_\tau)&\approx&
\left(\sin^22\theta_{\mbox{atm}}-\sin^22\theta_{13}\sin^2\theta_{23}\right)
\sin^2\frac{\Delta m^2_{\mbox{atm}}L}{4E}\cr
P_{\mbox{vac}}(\nu_\tau\longrightarrow \nu_e)&\approx&
\sin^22\theta_{13}\cos^2\theta_{23}\sin^2\frac{\Delta m^2_{\mbox{atm}}L}{4E}\cr
P_{\mbox{vac}}(\nu_e\longrightarrow \nu_x)&\approx&
\sin^22\theta_{13}\sin^2\frac{\Delta m^2_{\mbox{atm}}L}{4E}\cr
P_{\mbox{vac}}(\nu_\mu\longrightarrow \nu_x)&\approx&
\sin^22\theta_{\mbox{atm}}\sin^2\frac{\Delta m^2_{\mbox{atm}}L}{4E}
\label{eq:atm}
\end{eqnarray}
while the solar $\nu_e$ survival probability is
\begin{equation}
P_{\mbox{vac}}(\nu_e\longrightarrow \nu_e)\approx\cos^4\theta_{13}
\left(1-\sin^22\theta_{\mbox{sol}}\sin^2\frac{\Delta m^2_{\mbox{sol}}L}{4E}\right)
+\sin^4\theta_{13}
\label{eq:sol}
\end{equation}
with $\theta_{\mbox{sol}}=\theta_{12}$ and 
$\sin\theta_{\mbox{atm}}=\cos\theta_{13}\sin\theta_{23}$.
Therefore, atmospheric
neutrinos yield information about the last column of the mixing matrix
and solar neutrinos about the first row. If $\theta_{13}$ is small, then
both solar and atmospheric neutrinos can be analyzed using a two-neutrino
model with effective solar and atmospheric mixing angles.

The characteristic $L/E$ dependence of the vacuum oscillation probability
is modified by the matter density along the flight path of the neutrino, since
the coherent scattering amplitude of $\nu_e$ with matter
is different from $\nu_\mu$ or $\nu_\tau$, and sterile
neutrinos don't undergo such coherent scattering at all. 
To take into account these matter effects, the potential $V$ can be defined.
The matter potential is the same for $\nu_\mu$ and $\nu_\tau$,
but differs for
$\nu_e$ (or a sterile flavor). The same two-neutrino formulas may be used,
with the substitution of
\begin{eqnarray}
\left(\Delta m^2_{\mbox{mat}}\right)^2&=&\left(\Delta m^2\right)^2\times
\left[\left(\frac{2E\Delta V}{\Delta m^2}+\cos2\theta\right)^2+\sin^22\theta\right]\cr
\sin^22\theta_{\mbox{mat}}&=&\sin^22\theta\frac{\Delta m^2}{\Delta m^2_{\mbox{mat}}}
\end{eqnarray}
for $\Delta m^2$ and $\sin^22\theta$,
where $\Delta V$ is the potential difference. While the vacuum conversion
probabilities are invariant under $\theta\longrightarrow\pi/2-\theta$ (that
is $\sin2\theta\longrightarrow\sin2\theta$) the cos term changes sign
($\cos2\theta\longrightarrow-\cos2\theta$) and the effective $\Delta m^2$ and
mixing changes. Therefore the ``mirror symmetry'' around maximal mixing
($\theta=\pi/4$) is broken by the matter effects.
Depending on the sign of $\Delta V$, a resonance can occur~\cite{msw}
which leads to a large conversion probability, even if the vacuum mixing
is small.

\section{Atmospheric Neutrinos}
Atmospheric neutrinos originate when primary cosmic rays strike the
earth's atmosphere and produce a shower of pions, which decay into
muons and $\nu_\mu$'s. Each muon decay produces a $\nu_e$ as
well as a $\nu_\mu$. Super-Kamiokande (SK) is a cylindrical
50,000 ton water Cherenkov detector. An optical barrier separates
an ``inner'' concentric cylinder from the outer ``anti-counter''.
The inner cylinder (32,000 tons) is viewed by 11,146 inward-facing 
20'' photomultiplier tubes, the outer cylinder by 1885 outward-facing
8'' photomultiplier tubes. SK uses the anti-counter
(which surrounds the inner detector) to define ``fully contained events''
(no entering or exiting charged particles), ``partially contained
events'' (only exiting charged particles) and ``entering events''.
In each case, the direction of a charged particle is reconstructed using
the directionality of the Cherenkov light.

Cherenkov rings are predominantly produced by electrons and muons.
The Cherenkov ring produced by electrons typically has a ``fuzzy''
outer edge due to electromagnetic showers. A muon, on the other hand,
does not shower as much, and therefore its Cherenkov ring
has a ``sharp'' outer edge. This allows the construction of a
particle identification likelihood (PID)
which distinguishes ``$e$-like'' from ``$\mu$-like'' events.
The energy of fully contained events is measured using the amount
of produced Cherenkov light. The sample is split into a ``sub-GeV''
and a ``multi-GeV'' sample. If the brightest ring in an event containing
several Cherenkov rings is $\mu$-like, the event is most likely due to
a $\nu_\mu$ charged-current interaction. Since the decay photons
of neutral pions produce electrons by Compton scattering, a neutral-current
enhanced multi-ring sample can be defined using the PID and the $\pi^0$
invariant mass.

\begin{figure}[bt]
\noindent(a)\hspace*{7.5cm}(b)

\vspace*{-0.2cm}
\epsfxsize 8cm\epsfbox{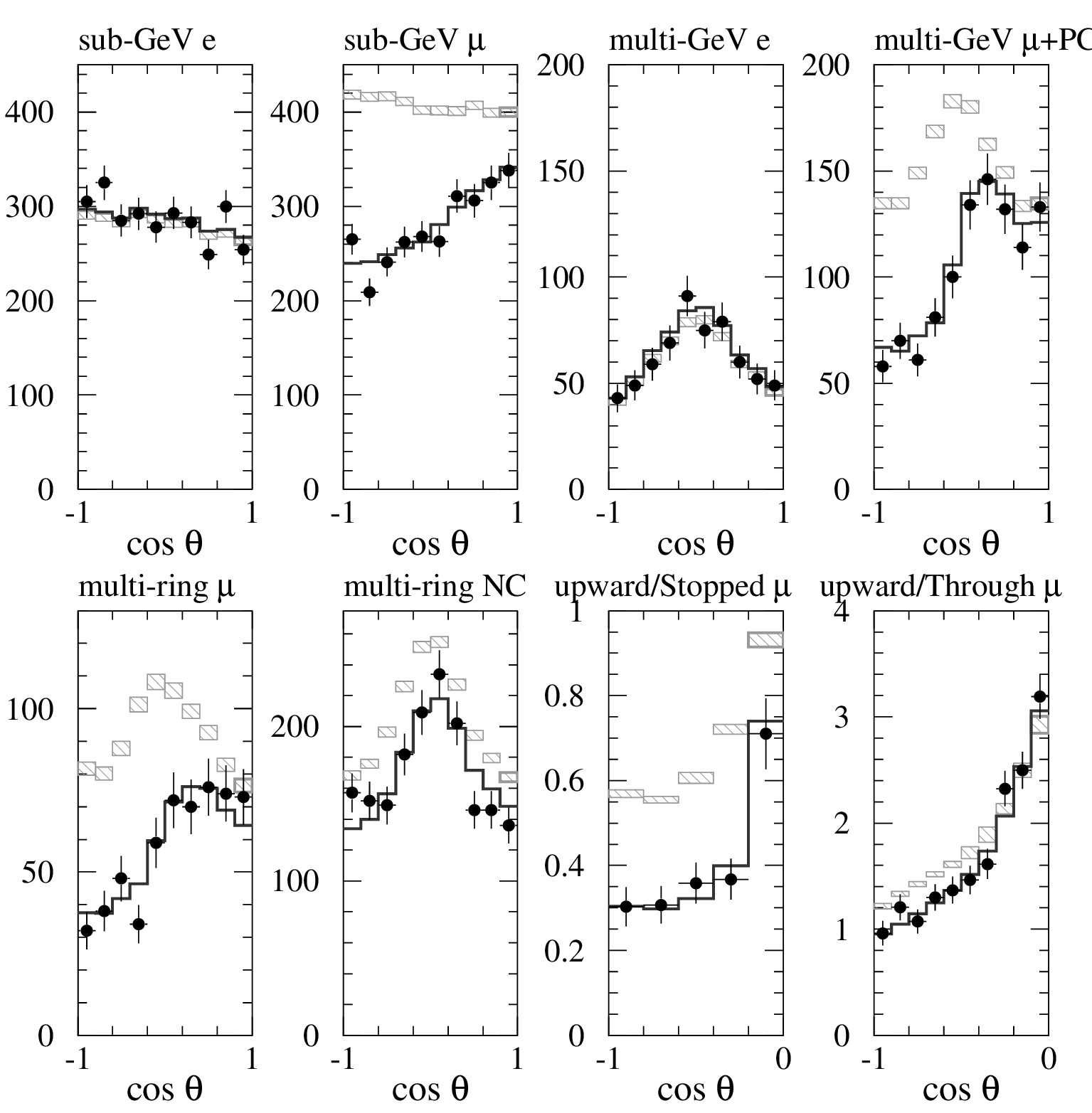}
\epsfxsize 8cm\epsfbox{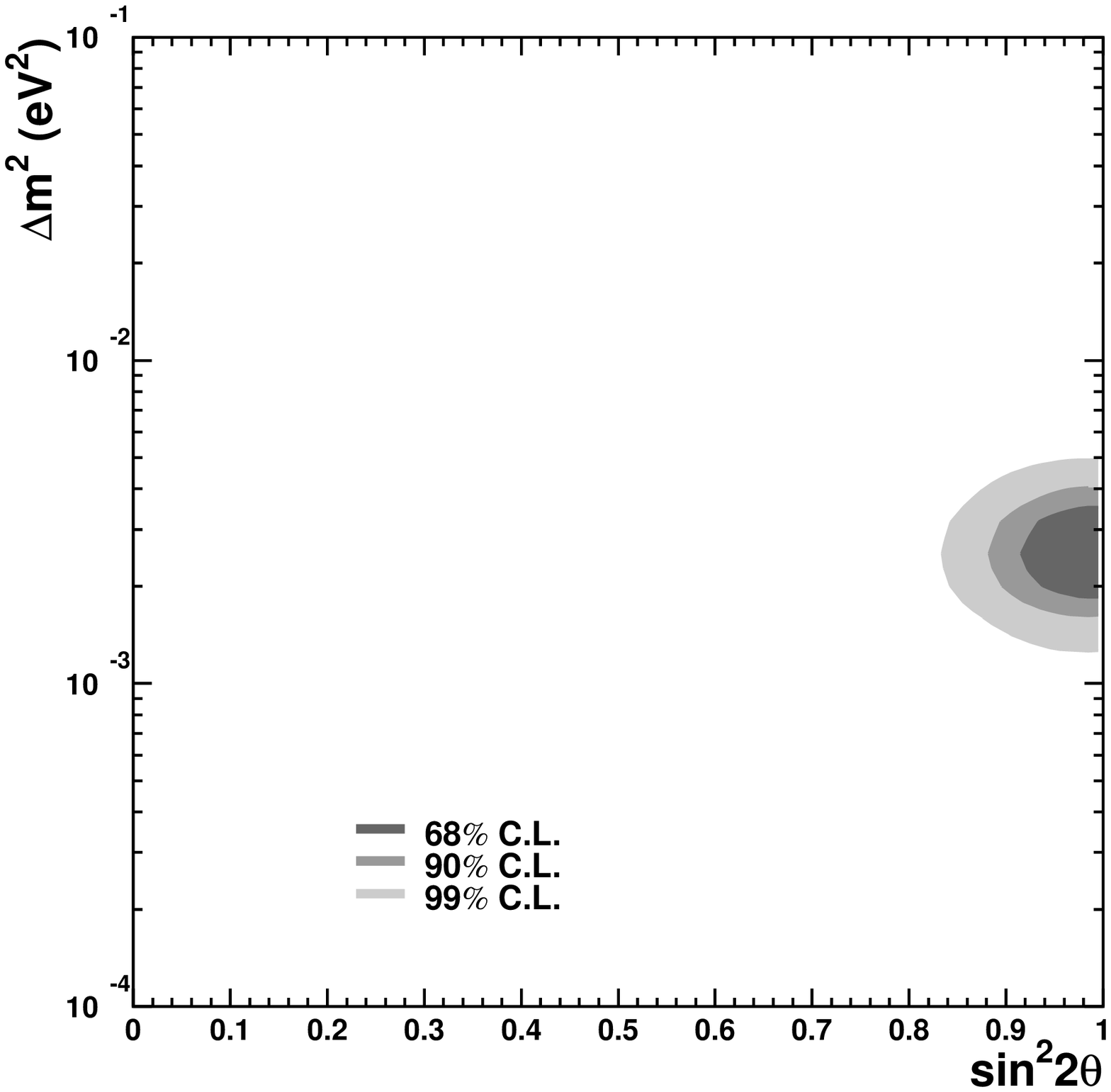}
\caption{SK Atmospheric Neutrino Data and Atmospheric Neutrino Oscillation
Parameters.
(a) The upper four panels show the zenith angle distribution of single
ring events for fully contained $e$-like rings (panel one and three),
fully contained $\mu$-like rings (panel two) and higher energy
$\mu$-like rings (panel four). In the lower row,
panel one (brightest ring is $\mu$-like)
and two (brightest ring is $e$-like: neutral-current enhanced sample)
contain the zenith angle distributions of the multi-ring
samples. The zenith angle distributions of $\nu_\mu$ interacting
in the rock below SK are given in panel three (lower energy) and
four (higher energy). The predictions of the atmospheric neutrino
Monte Carlo calculation are overlaid (light-gray hatched areas).
The best oscillation fit is indicated by the dark-gray solid line.
(b) Allowed oscillation parameters for a two-neutrino fit assuming pure
$\nu_\mu\leftrightarrow\nu_\tau$ oscillations.}
\label{fig:numunutau}
\end{figure}

Partially contained events are almost always due to higher energy
muons. The energy of those muons can only be estimated statistically 
using Monte Carlo distributions. Upward-going entering events are
due to the interaction of $\nu_\mu$ in the rock below SK. If the
``upward-going muons'' stop inside the detector (``upward/stopped''),
then the average energy of the producing $\nu_\mu$'s is lower than
for upward-going muons which enter and exit the detector 
(``upward/through'').

The direction of low
energy leptons yield little information about the neutrino's direction
(and in turn its flight distance). However, the zenith angle $\theta$ of
high energy leptons is correlated with the neutrino flight distance $L$.
SK has accumulated 1289 live days of atmospheric neutrino data
using a fiducial volume of 22,500 tons.
The zenith distributions of electrons (see figure~\ref{fig:numunutau})
show good agreement with the Monte Carlo based on the atmospheric neutrino
flux calculation. However, there are significantly less muons than expected.
Downward-going ($\cos\theta=1$) atmospheric neutrino-induced
(multi-GeV) $\mu$-like events agree with the Monte Carlo,
upward-going events show the strongest deficit. A similar distortion
is seen in the multi-ring $\mu$-like sample, but not in the multi-ring
neutral-current enhanced sample.
Upward-going through-going muons agree with Monte Carlo 
in the horizontal direction, but are suppressed at $\cos\theta=-1$.
There are less upward-going stopped muons than expected.
As seen in figure~\ref{fig:numunutau} (a) oscillations of $\nu_\mu$'s
into $\nu_\tau$'s are able to explain all the zenith angle distributions. 
The same figure also gives allowed ranges for neutrino mixing and
$\Delta m^2$. The data prefer maximal mixing and 
$\Delta m^2\approx$2--3$\cdot10^{-3}$eV$^2$.

\begin{figure}[bt]
\noindent(a)\hspace*{7.5cm}(b)

\vspace*{-0.2cm}
\epsfxsize 8cm\epsfbox{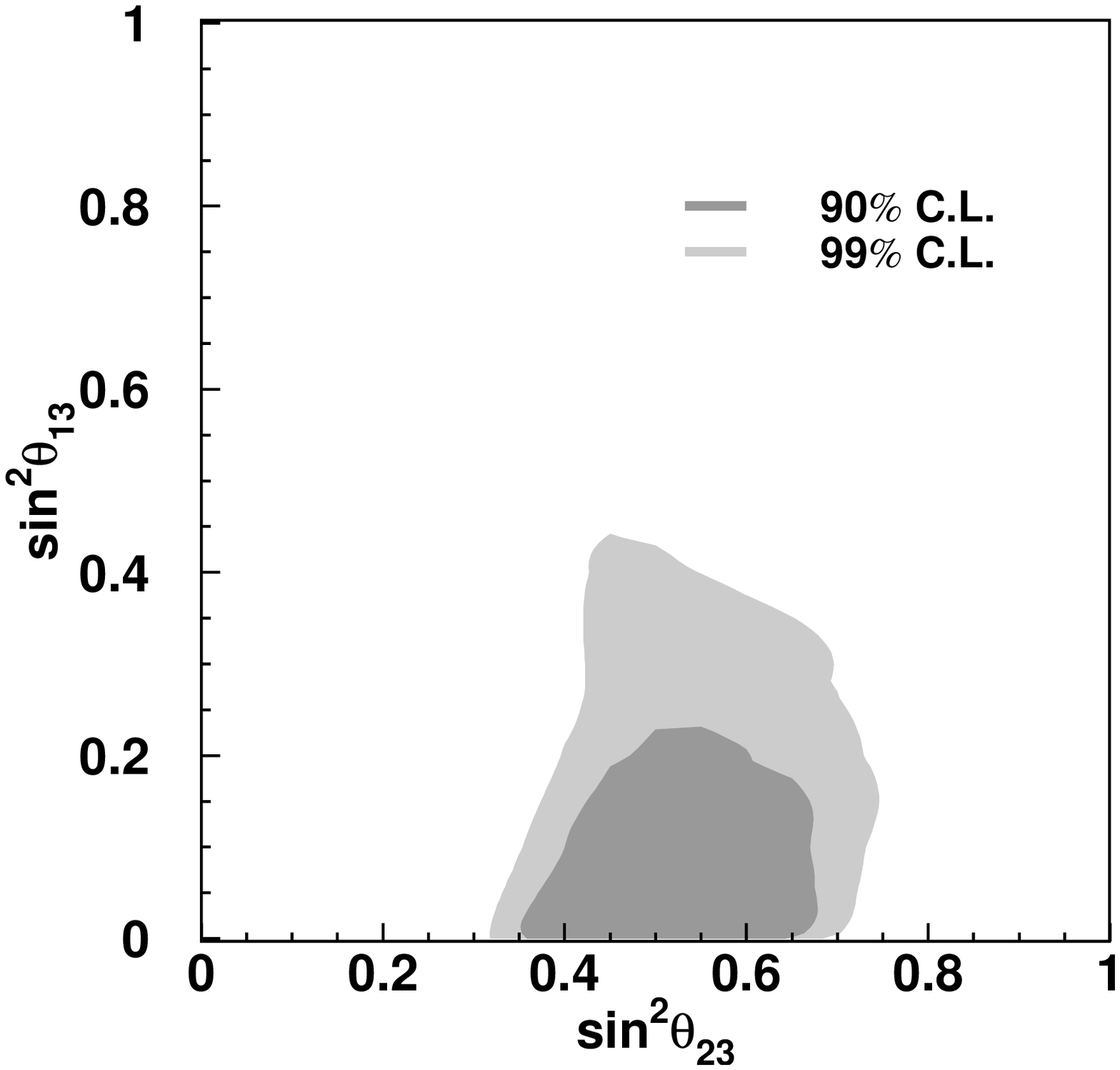}\epsfxsize 8cm\epsfbox{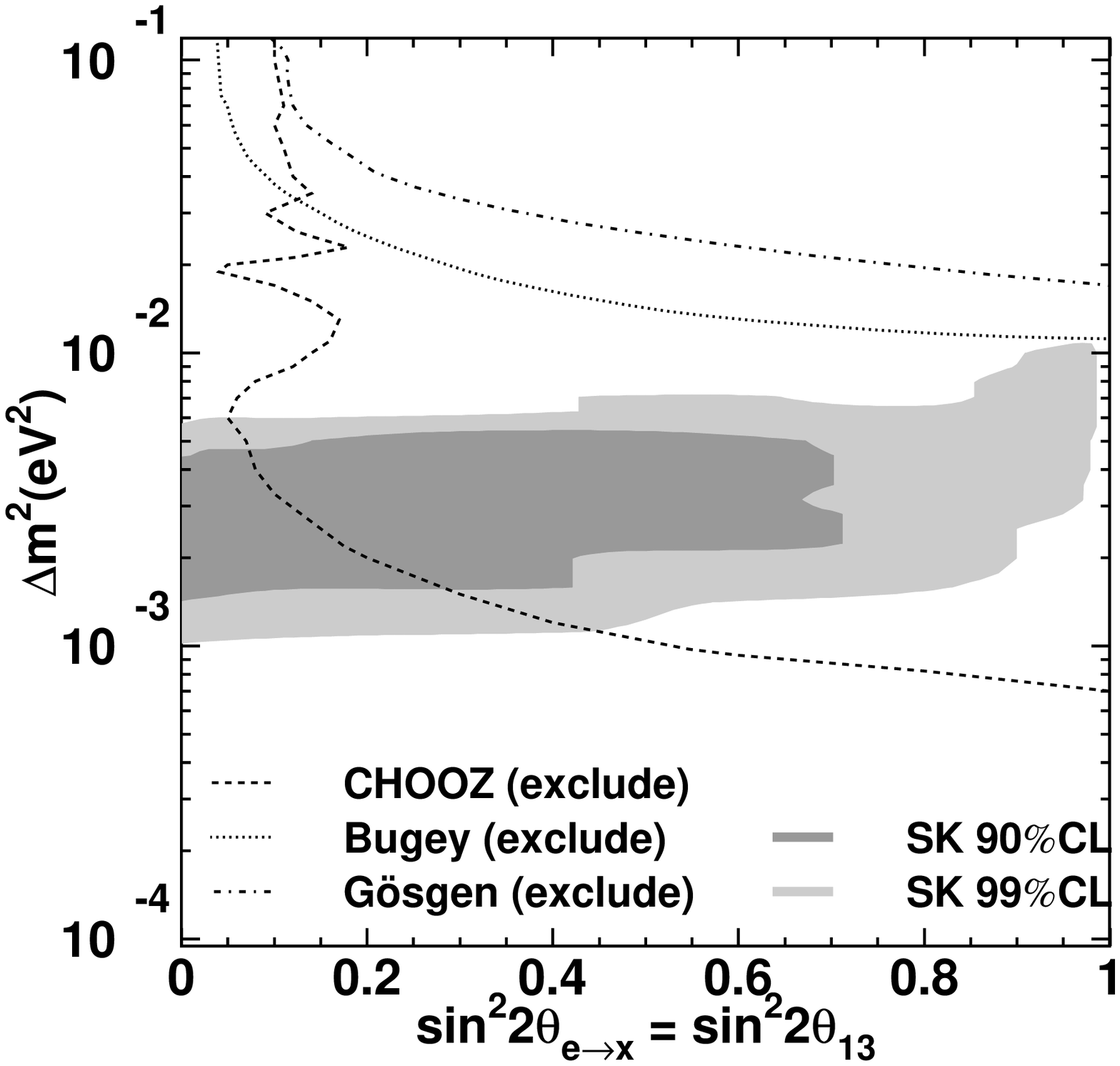}
\caption{Neutrino Mixing with Three Parameters.
(a) Allowed area of $\sin^2\theta_{13}$ and $\sin^2\theta_{23}$.
At 90\% C.L. SK data allows up to about 40\% of $e$-type
content of the atmospheric oscillation.
(b) Allowed range of the atmospheric $\nu_e$ disappearance probability
compared to the limits for $\overline{\nu_e}$ obtained from various
reactor experiments. The most stringent limit comes from the
CHOOZ experiment.}
\label{fig:oscmix}
\end{figure}

Due to the lack of an $e$-like appearance signal,
$\nu_\mu\leftrightarrow\nu_e$ oscillations are disfavored compared
to $\nu_\mu\leftrightarrow\nu_\tau$. However, any $e$-like appearance
due to $\nu_\mu\rightarrow\nu_e$ is unfortunately 
washed out by a disappearance due
to $\nu_e\rightarrow\nu_\mu$, so the SK data allow a substantial
``$\nu_e$'' component (about 40\% at 90\% C.L.) in the $\nu_\mu$ oscillations.
Figure~\ref{fig:oscmix} shows the allowed values for both
$\theta_{13}$ and $\theta_{23}$
(according to equation~(\ref{eq:atm})
$\theta_{12}$ plays no role for atmospheric neutrino oscillations)
in a three-neutrino oscillation analysis. Maximal mixing for
$\theta_{23}$ and zero mixing for $\theta_{13}$ is preferred.
The same figure also gives allowed ranges for $\Delta m^2_{\mbox{atm}}$
and the ``disappearance probability'' $\sin^22\theta_{13}$. More
stringent limits on this probability come from the CHOOZ reactor
experiment~\cite{chooz}.

\begin{figure}[bt]
\noindent(a)\hspace*{7.5cm}(b)

\vspace*{-0.8cm}
\epsfxsize 8cm\epsfbox{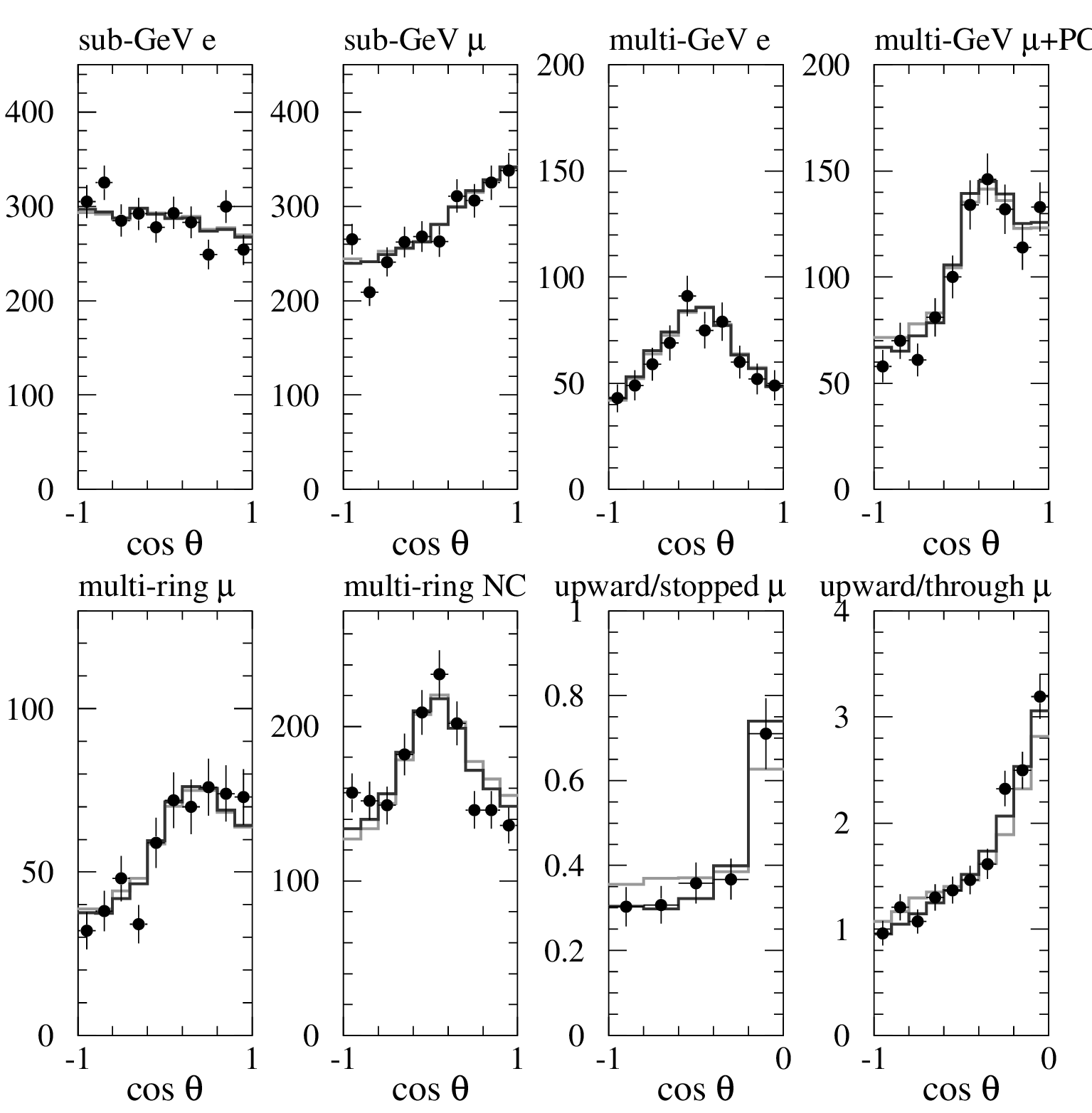}\epsfxsize 8cm\epsfbox{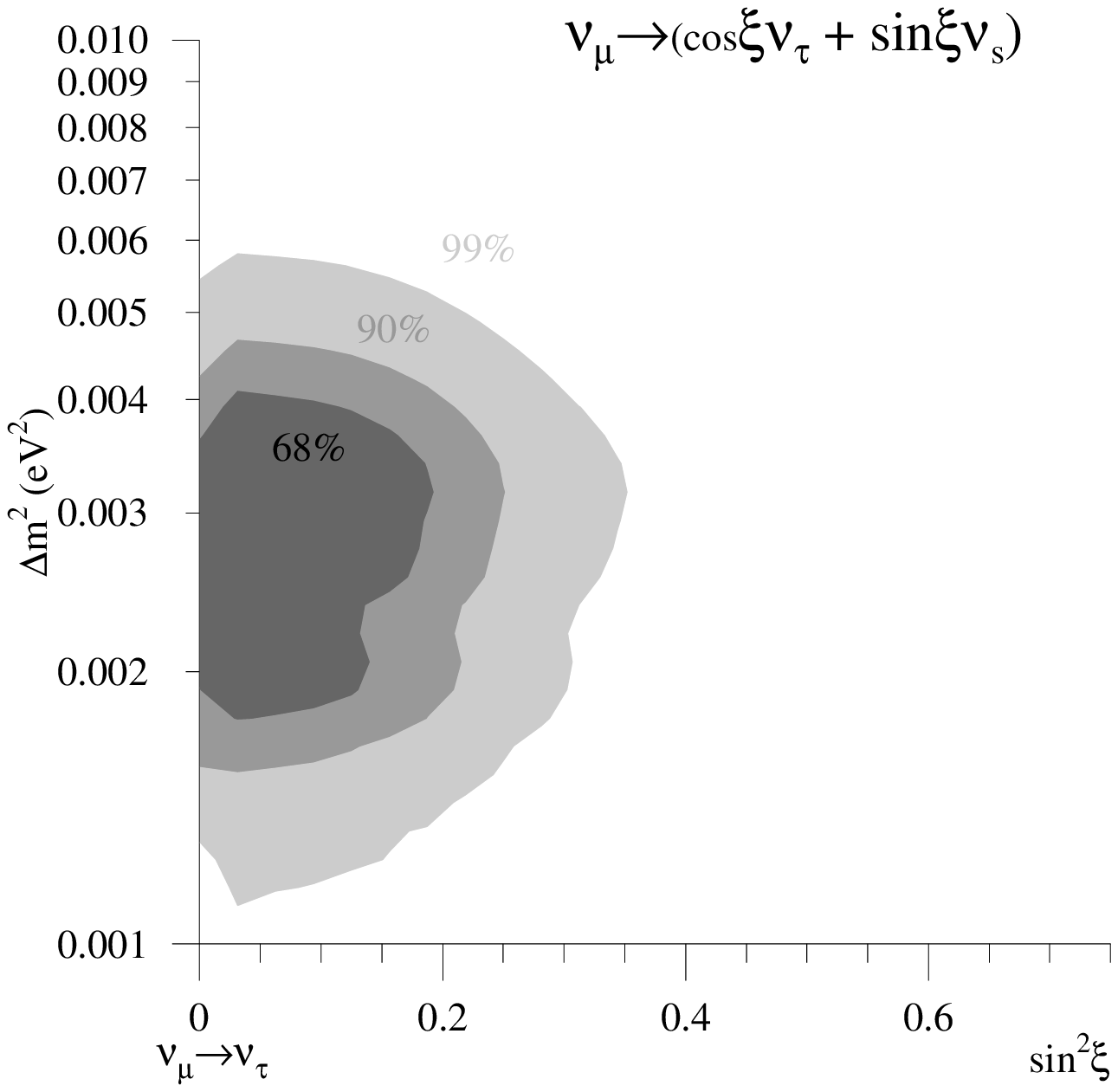}
\caption{Limit on Sterile Content of Atmospheric Neutrino Oscillations.
(a) Pure $\nu_\mu\leftrightarrow\nu_\tau$ (solid black line) fits
the upward-going $\mu$ zenith distribution better than
pure $\nu_\mu\leftrightarrow\nu_s$ (solid gray line) because of matter
effects. For the same reason, there is also a small difference in the 
high energy $\mu$-like events.
The neutral-current enhanced multi-ring sample shows no hint for a
disappearance of active flavors and therefore favors
pure $\nu_\mu\leftrightarrow\nu_\tau$.
(b) The limit on the sterile content is $\sin^2\xi<0.25$ at 90\% C.L.}
\label{fig:numunus}
\end{figure}

Four-neutrino oscillations are investigated for the
case of $\theta_{13}=0$ and a
$\Delta m^2$ hierarchy 
($\Delta m^2_{\mbox{LSND}}\approx1$eV$^2\gg\Delta m^2_{\mbox{atm}}\approx10^{-3}$eV$^2\gg\Delta m^2_{\mbox{sol}}\approx10^{-4}$eV$^2$).
The ``content'' of the fourth flavor (sterile neutrino)
in atmospheric neutrino oscillations can be
parameterized by $\sin^2\xi$:
$\nu_\mu\rightarrow(\cos\xi\nu_\tau+\sin\xi\nu_s)$.
Figure~\ref{fig:numunus} compares the best
pure $\nu_\mu\rightarrow\nu_\tau$ fit with the best
pure $\nu_\mu\rightarrow\nu_s$ fit and gives the allowed ranges for 
$\Delta m^2_{\mbox{atm}}$ and $\sin^2\xi$.
In case of purely sterile oscillations, matter effects reduce the
conversion probability for high energy neutrinos which affects the
upward-going muon samples, the partially contained events, and the
multi-GeV $\mu$-like events. The neutral-current enhanced multi-ring
zenith angle distribution becomes up/down asymmetric, since sterile neutrinos
don't interact via neutral currents. In every case, data disfavors
the purely sterile case. Data likes best zero sterile content
At 90\% C.L., the maximum allowed sterile content is 25\%. The range
of $\Delta m^2$ is consistent with the other fits.

Other hypotheses to explain the SK zenith angle distributions are
investigated as well. The conversion probability is parameterized as follows:
(i) $\sin^22\theta\sin^2(\alpha L\times E)$ (``LxE''),
(ii) $\sin^4\theta+\cos^4\theta\left(1-e^{-\alpha L/E}\right)$
(``$\nu_\mu$ decay'') and
(iii)  $\left(\sin^2\theta+\cos^2\theta e^{-\alpha L/E}\right)^2$
(``$\nu_\mu$ decay to $\nu_s$''). Table~\ref{tab:otherhyp}
gives a summary of all investigated hypotheses.
$\nu_\mu\rightarrow\nu_\tau$ is strongly favored by the data.

\begin{table}[htb]
\caption{Comparison of Several Flavor Conversion Hypotheses.
$\nu_\mu\rightarrow\nu_\tau$ is strongly favored by the data.}
\label{tab:otherhyp}
\vspace{0.4cm}
\begin{center}
\begin{tabular}{|c|cc|cc|}
\hline
Mode & \multicolumn{2}{c|}{best fit parameters} &
$\Delta\chi^2$ & $\# \sigma$ \cr \hline
$\nu_\mu\rightarrow\nu_\tau$      & $\sin^22\theta=1.00$ &
$\Delta m^2=2.5\cdot10^{-3}$eV$^2$&  0.0 & 0.0 \cr
$\nu_\mu\rightarrow\nu_e$         & $\sin^22\theta=0.97$ &
$\Delta m^2=5\cdot10^{-3}$eV$^2$  &  79.3 & 8.9 \cr
$\nu_\mu\rightarrow\nu_s$         & $\sin^22\theta=0.96$ &
$\Delta m^2=3.6\cdot10^{-3}$eV$^2$& 19.0 & 4.4 \cr
LxE                               & $\sin^22\theta=0.90$ &
$\alpha=5.3\cdot10^{-4}$/(GeVkm)  & 67.1 & 8.2 \cr
$\nu_\mu$ decay                   & $\cos^2\theta=0.47$  &
$\alpha=3.0\cdot10^{-3}$GeV/km    & 81.1 & 9.0 \cr
$\nu_\mu$ decay to $\nu_s$        & $\cos^2\theta=0.33$  &
$\alpha=1.1\cdot10^{-2}$GeV/km    & 14.1 & 3.8 \cr
\hline
\end{tabular}
\end{center}
\end{table}

\begin{figure}[bt]
\noindent(a)\hspace*{7.5cm}(b)

\vspace*{-0.2cm}
\epsfxsize 8cm\epsfbox{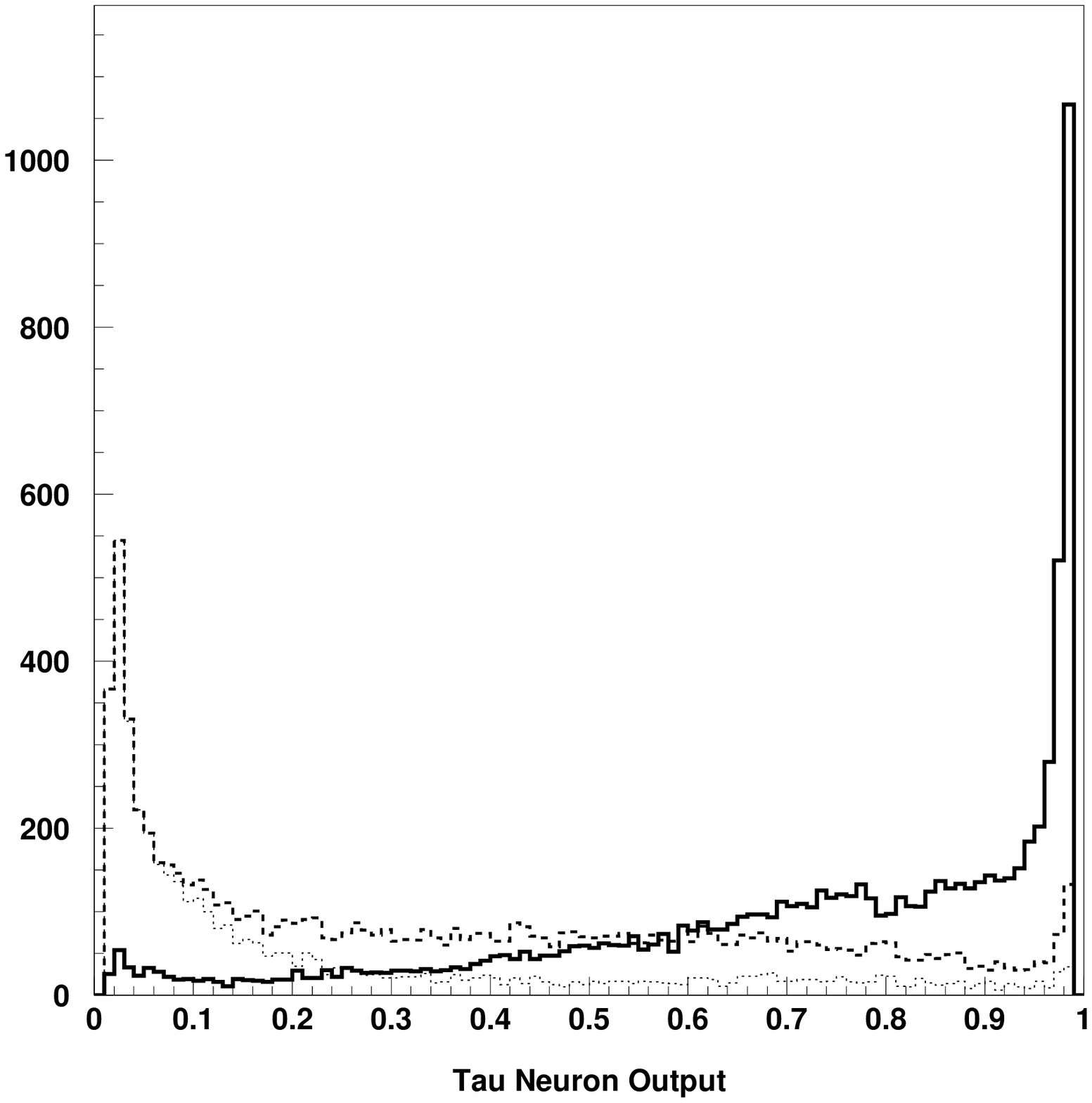}\epsfxsize 8cm\epsfbox{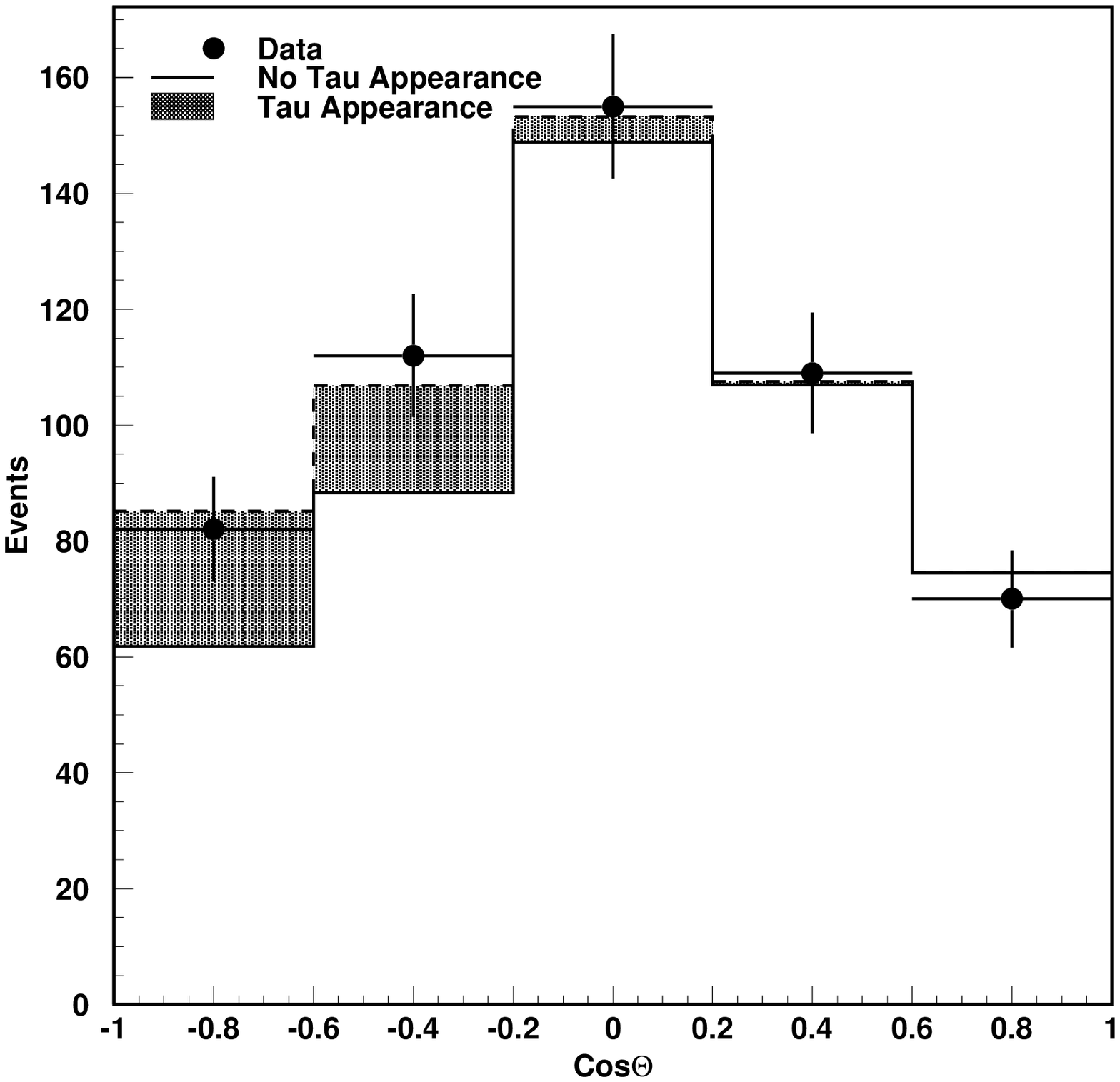}
\caption{Search for Tau Appearance.
(a) Neural net output for atmospheric $\nu_\mu$ charged-current
interaction Monte Carlo (dotted line), all atmospheric $\nu$ Monte Carlo
without $\nu_\tau$ (dashed line) and tau Monte Carlo (solid line).
Since there is only a small number of tau events expected in the SK
data, the analysis is limited by background from other atmospheric
$\nu$-induced events.
(b) Fit to the zenith angle distribution of the tau-enriched sample.
Tau events are only expected in the upward direction. Data favors
tau appearance at about the $2\sigma$ level.}
\label{fig:tau}
\end{figure}

If atmospheric $\nu_\mu$ indeed oscillate into $\nu_\tau$,
then about 80 $\nu_\tau$ charged-current events are expected
in the SK data set. The high threshold for this reaction makes
it difficult to isolate these events. At this energy, atmospheric
neutrinos typically produce a spray of particles, and ring-counting
becomes very difficult. One analysis abandons ring-counting altogether
and reconstructs the amount of ``energy flow'' as a function of direction
based on the angular light distribution. From the energy flow,
jet variables are
formed. A $\tau$ likelihood function
combines such variables to discriminate between
$\tau$ events and other atmospheric neutrino interactions.
Another analysis constructs a similar $\tau$ likelihood from
variables based on the conventional event reconstruction (ring-counting)
A third analysis utilizes a neural net to combine similar
ring-counting variables. Figure~\ref{fig:tau} (a)
explains the output of the ``tau neuron'' when tested
on Monte Carlo. In spite of the good separation between tau and
non-tau events, the analysis (like the other two) is limited by
background, since there are much fewer tau events than non-tau events.
The same figure shows the zenith angle distribution of the tau-enriched
sample (using the tau neuron). An excess of about
two sigma (see also table~\ref{tab:tauresults}) is observed
for upward-going events.

\begin{table}[hbt]
\caption{Results of Three $\tau$ Appearance Searches.}
\label{tab:tauresults}
\vspace{0.4cm}
\begin{center}
\begin{tabular}{|c|cc|cc|}
\hline
Analysis & Number of $\tau$-Events in Fit & Efficiency &
Observed, & Expected Signif. \cr
\hline
Energy-Flow   & 79$^{+44}_{-40}$(stat+syst)         & 32\% &
1.8$\sigma$ & 1.9$\sigma$ \cr
Ring-Counting & 66$\pm$41(stat)$^{+25}_{-18}$(syst) & 43\% &
1.5$\sigma$ & 2.0$\sigma$ \cr
Neural Net    & 92$\pm$35(stat)$^{+18}_{-23}$(syst) & 51\% &
2.2$\sigma$ & 2.0$\sigma$ \cr
\hline
\end{tabular}
\end{center}
\end{table}

\section{Solar Neutrinos}

SK observes solar neutrinos via elastic neutrino-electron scattering.
Only solar neutrinos with fairly high energy (several MeV) can be
detected. These neutrinos originate from the $\beta^+$-decay of $^8$B
or the $^3$He-proton ({\it hep}) fusion reaction.
Like other experiments SK found a deficit of solar neutrinos;
all such deficits are explained by solar neutrino oscillations.
Since event times are recorded and the event energy is reconstructed,
SK can search for time variations and an energy dependence of the
rate deficit.

\begin{figure}
\epsfxsize 15cm
\epsfbox{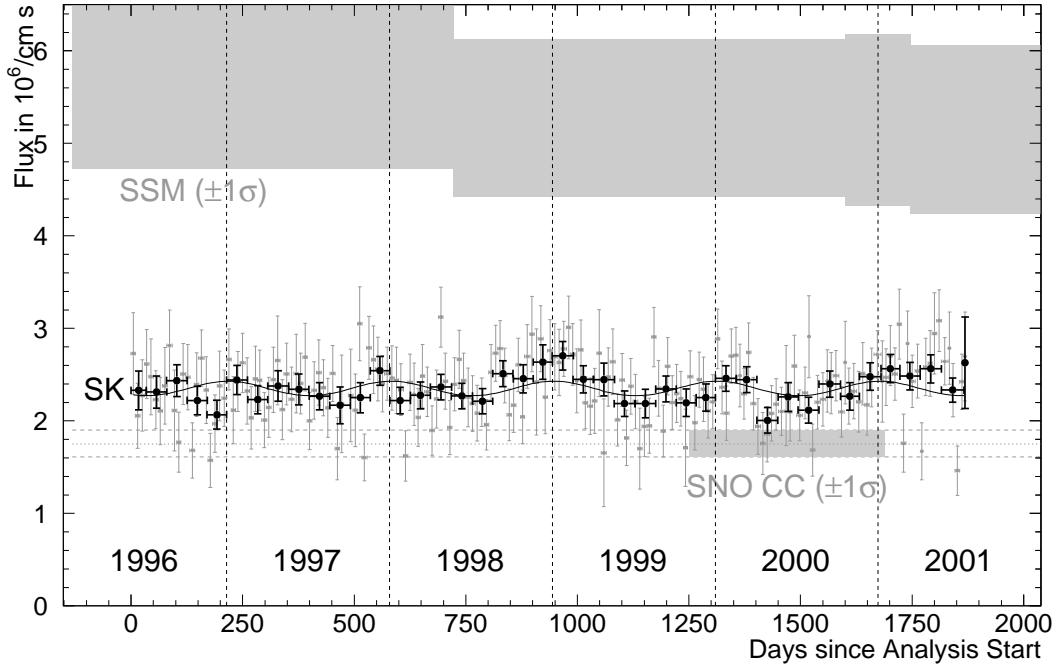}
\caption{Time Variation of the Neutrino Flux Inferred from
the Elastic Scattering Rate.
The black points show the elastic scattering rate in SK during
1.5 months. The gray points extend over a
time period of 10 days. Each black point is significantly below
the SSM prediction and above the SNO charged-current rate.
The solid line shows the 7\% flux variation which is expected from the
3.5\% change in the distance between the sun and the earth.
}
\label{fig:timevar}
\end{figure}

The solar neutrino flux is inferred from the SK interaction rate
assuming only $\nu_e$'s.
The large solar neutrino event sample collected by SK during 1496
live days inside a fiducial volume of 22,500 tons
enables a precise study of the time-dependence of this flux.
The gray data points in figure~\ref{fig:timevar} contain only 10 days of data.
The flux measured in each of these 185 time bins is significantly
smaller than the SSM prediction. Also shown are the larger
1.5 month time bins
(black data points). At first glance, the flux looks constant. A more
careful study reveals an indication for a periodic change with
a one year period. Since the sun is a neutrino point source,
a 7\% change of that kind is expected by the
3.5\% change in distance between the sun and the earth (see solid
black line of figure~\ref{fig:timevar}). If the 1.5 month data points
are combined into 8 ``seasonal'' bins, a $\chi^2$ test to this
7\% variation yields $\chi^2=4.7$ (69\% C.L.) compared to
$\chi^2=10.3$ (17\% C.L.) for a constant flux. The difference in
$\chi^2$ corresponds to about 2.5 sigma. There is
no hint of other long-term time variations 
(e.g. arising from the oscillation phase).

The charged-current interaction rate measured
by the SNO experiment explores the solar neutrino flux in a similar
energy range and can therefore be compared directly.
The flux measured by SK in each of the 42 (1.5 month)
time bins is significantly larger
then the flux measured by SNO. Solar neutrino oscillations of $\nu_e$ into
other, active flavors predict such an excess of the elastic scattering
rate over a charged-current measurement;
that excess is due to a larger neutral-current amplitude in the
elastic-scattering process. The discrepancy between the SK and the SNO
neutrino flux measurement is therefore interpreted as evidence for
the presence of other active flavors than $\nu_e$
in the solar neutrino flux.

\begin{figure}
\epsfxsize 15cm
\epsfbox{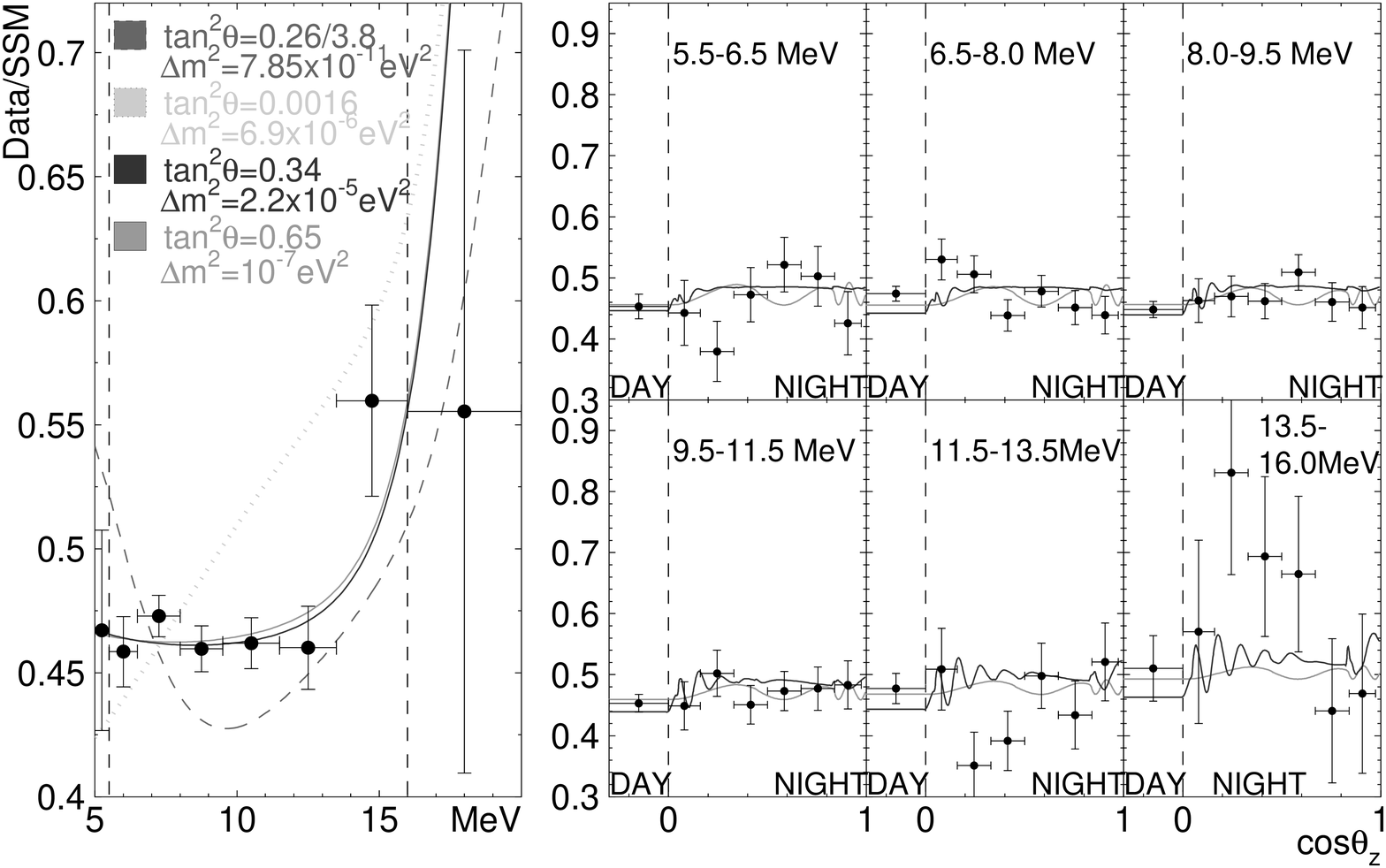}
\caption{Zenith Spectrum and Four Oscillation Predictions.
SMA (dotted line) and VAC (dashed line) solutions are disfavored
by spectral distortion data, LOW (light gray) solutions by the absence
of solar zenith angle variation. The LMA solutions (dark gray) fit best.
\label{fig:zenspec}}
\end{figure}

Depending on the solar $\Delta m^2$,
solar neutrino oscillations can be driven by the oscillation
phase or by matter effects. Strong spectral distortions are
expected in the former case. In the latter case, spectral distortions
can still occur as a consequence of a (energy-dependent) resonance caused by
the sun's matter density. The earth's matter density can also
affect the conversion probability resulting in most cases in a
``regeneration'' of ``disappeared'' $\nu_e$'s during the night.
SK searches for both effects with the ``zenith angle spectrum''
(see figure~\ref{fig:zenspec}): The data is broken into eight
energy bins (according to the SK energy resolution). Between
5.5 and 16 MeV of reconstructed energy, SK has collected a sufficiently
large number of solar neutrino candidate events to subdivide each energy
bins in seven solar zenith angle ($\theta_z$) bins. In each bin,
the ratio of the observed event rate and the rate expected by the
SSM is shown.
Neither spectral distortion
nor a regeneration effect is evident in the data.

\begin{figure}
\epsfxsize 15cm
\epsfbox{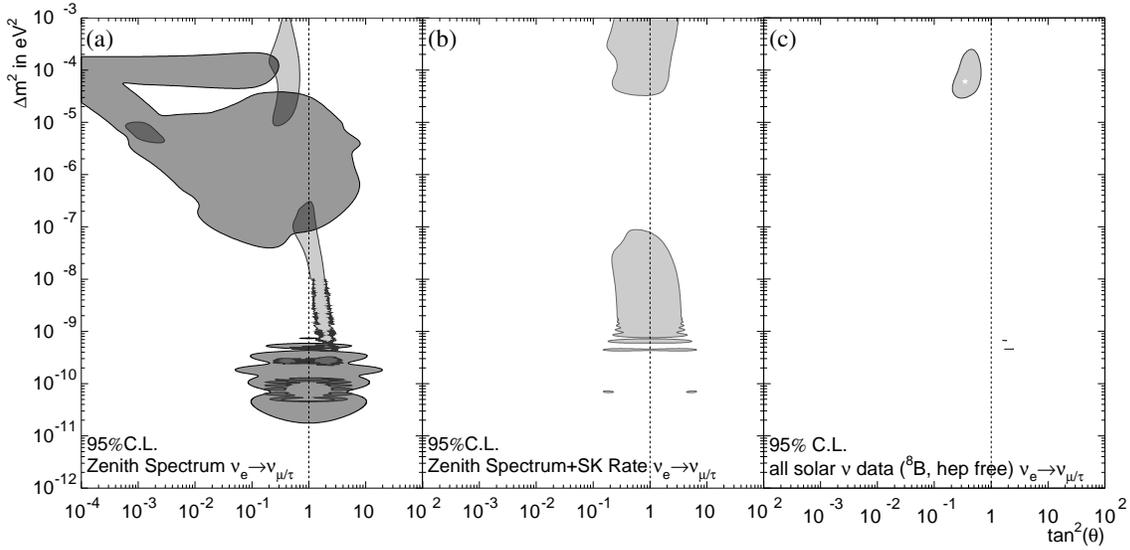}
\caption{Solar Neutrino Oscillation Parameters.
(a) The gray area is excluded at 95\% C.L. by the shape of the SK zenith angle
spectrum, the light-gray area is allowed by data from Gallex/GNO, SAGE,
Homestake and SNO. The overlap of both is shaded dark-gray.
(b) The light-gray area is allowed by the shape of the SK zenith angle
spectrum and the SK interaction rate using the SSM prediction
of the $^8$B flux.
(c) The light-gray area is allowed using all solar neutrino data. The
SSM predictions of the $^8$B and {\em hep} flux are not used.
The best fit (located in the LMA solution) is marked by the white
asterisk.
\label{fig:solosc}}
\end{figure}

For the oscillation analysis~\cite{osc,lma} of this
zenith angle spectrum (and other solar
data), $\theta_{13}$ is set to zero in
equations~(\ref{eq:sol}). This
is justified by the lack of evidence for a positive $\theta_{13}$
in either atmospheric neutrino data from SK or reactor neutrino data from
CHOOZ. Analysis of the data from the other SK experiments reveal several
distinct allowed areas shown in light-gray in figure~\ref{fig:solosc} (a)
which are referred to as ``solutions'' to the solar neutrino
problem. The small mixing angle solution (SMA) is the only one with
a small solar mixing angle. There are three large angle solutions.
The large mixing angle solution (LMA) has the largest $\Delta m^2$
and the vacuum solution (VAC) the smallest. The lower part of
the region extended between $10^{-9}$eV$^2$ and  $10^{-7}$eV$^2$
is called quasi-VAC solution, the upper part is the LOW solution.
Figure~\ref{fig:zenspec} shows the typical spectral and solar zenith
angle behavior of the main four solutions. The absence of spectral
distortions excludes VAC and SMA, the absence of the nightly
``regeneration shape'' required by the LOW solution disfavors this solution
as well. Figure~\ref{fig:solosc} (a) shows the excluded regions of parameter
space due to only the shape of the zenith angle spectrum (gray
areas). Only LMA and quasi-VAC solutions remain allowed. Combining
the shape of the zenith angle spectrum with the SK rate measurement,
only the two allowed regions in figure~\ref{fig:solosc} (b)
remain (LMA and quasi-VAC). SK solar neutrino data by itself requires
large mixing and prefers maximal mixing. Stronger constrains are obtained
when the data of the other solar neutrino experiments are used
in addition to SK data. The quasi-VAC solution
is rejected by the rates of all experiments.
Combining~\cite{lma} SK zenith angle spectrum and rate with the
rates of all other solar neutrino
experiments, SK found for the first time a {\it unique} solution,
the LMA solution, at $\approx$95\% C.L. Figure~\ref{fig:solosc} (c) shows
the allowed range in mixing and $\Delta m^2$. The mixing is large
but not maximal and $\Delta m^2$ is somewhat smaller then $10^{-4}$eV$^2$.
Since SK spectral data limits high energy solar neutrino spectral
distortions to be small, the combination of the SK rate and the
charged-current rate of either the Homestake experiment or SNO
forbids a purely sterile solar neutrino oscillation solution.

\begin{figure}[hbt]
\vspace*{4cm}\hspace*{0.4cm}$
U\approx\pmatrix{
 \frac{\sqrt{3}}{2}  &  \frac{1}{2}                & 0 \cr
-\frac{1}{2\sqrt{2}} &  \frac{\sqrt{3}}{2\sqrt{2}} & \frac{1}{\sqrt{2}}\cr
\frac{1}{2\sqrt{2}}  & -\frac{1}{2\sqrt{2}}        & \frac{1}{\sqrt{2}}
}
$

\vspace*{0.5cm}\hspace*{0.5cm}$
\theta_{12}\approx\frac{\pi}{6} \hspace*{0.1in}
\theta_{23}\approx\frac{\pi}{4} \hspace*{0.1in}
\theta_{13}\approx0
$

\vspace*{-4.5cm}\hspace*{5.2cm}
\epsfxsize 11cm
\epsfbox{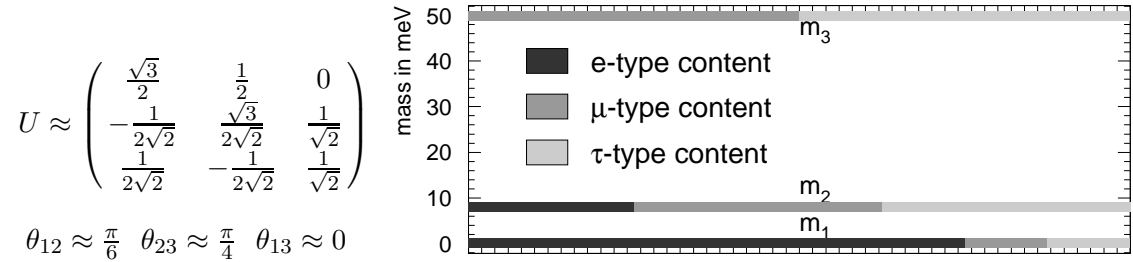}
\caption{A Possible Leptonic Mixing Matrix U and Neutrino Mass Spectrum.
The mixing matrix neglects CP violation and assumes that $\theta_{13}$
is indeed zero and the atmospheric mixing is maximal. The mass spectrum
additionally assumes, that $m_1$ is small enough to be negligible and
that the masses of the``solar pair'' are smaller than the third mass.
(hierarchical case).}
\label{fig:mass}
\end{figure}

\section{Neutrino Mixing and Masses}

Figure~\ref{fig:mass} summarizes and combines the current best knowledge
of neutrino masses and mixing from solar and atmospheric neutrino data.
There are no hints for sterile neutrinos in either atmospheric or
solar neutrino oscillations, so three neutrinos are sufficient to
explain both data sets. The mass spectrum of figure~\ref{fig:mass} is
for the hierarchical case, that is $m_1$ is much smaller than
the other masses, and the mass eigenstate pair responsible for solar
neutrino oscillations is $m_1$ and $m_2$. CP violating phases are
neglected. The displayed values of $\theta_{ij}$ are the best-fit
values from the atmospheric and the solar oscillation analyses.

\section*{Acknowledgments}
We gratefully acknowledge the cooperation of the Kamioka Mining and Smelting 
Company.
The Super-Kamiokande detector has been built and
operated from funding by the Japanese Ministry of Education, Culture,
Sports, Science and Technology, the U.S. Department of Energy, and the
U.S. National Science Foundation.

\end{document}